\documentclass[11pt]{article}
\usepackage{epsfig,dsfont}

\begin{document}
\sffamily

\thispagestyle{empty}
\vspace*{15mm}

\begin{center}

{\LARGE 
Chiral symmetry and spectral properties 
\vskip2mm
of the Dirac operator in G$_2$ Yang-Mills Theory
}
\vskip20mm
Julia Danzer, Christof Gattringer and Axel Maas
\vskip5mm
Institute for Physics, Universit\"atsplatz 5, Karl-Franzens University Graz \\
A-8010 Graz, Austria 
\end{center}
\vskip30mm

\begin{abstract}
We study spontaneous chiral symmetry breaking and the spectral properties 
of the staggered lattice Dirac operator using quenched gauge 
configurations for the 
exceptional group G$_2$, which has a trivial center. 
In particular we study the system below and above the finite
temperature transition and use the temporal boundary conditions of the 
fermions to probe the system. We evaluate several observables: The spectral
density at the origin, the spectral gap, the chiral condensate and the
recently proposed dual chiral condensate. We show that chiral 
symmetry is broken at
low temperatures and is restored at high temperatures at the 
thermodynamic phase transition. Concerning the role of the
boundary conditions we establish that in all respects the spectral 
quantities behave for G$_2$ in exactly the same way as for SU($N$),
when for the latter group the gauge ensemble above $T_c$ is restricted 
to the sector of configurations with real Polyakov loop.  
\end{abstract}

\setcounter{page}0
\newpage
\noindent
{\Large  Introductory remarks}
\vskip5mm
\noindent
Confinement of quarks and the breaking of chiral symmetry are two of the key
features of QCD. As the temperature is increased, QCD changes its behavior:
Quarks become deconfined and chiral symmetry is restored. It is a
long standing question whether the two phenomena and their changing behavior
at the QCD finite temperature phase transition are linked by some 
underlying mechanism. 

The gauge groups usually considered in Yang-Mills theories are SU($N$), in
particular SU(3). These groups have the non-trivial center groups 
$\mathds{Z}_N$ (and $\mathds{Z}_3$, respectively). 
It has been speculated that the center degrees of freedom play
an important role for confinement (see \cite{greensite} and references
therein), as well as for chiral symmetry breaking \cite{chirsimcenter}.

However, once dynamical fermions in the fundamental representation 
are introduced, the center is explicitly
broken, and confinement is no longer signaled by an infinitely rising
potential. Nonetheless, the chiral transition and the transition manifest
in residual center observables still coincide.

On the other hand, when the quarks are in the adjoint 
representation \cite{adjointtransition}, the
center symmetry is still intact at low temperatures, 
and the finite temperature transition
is signaled by a change of center-sensitive observables, like the Polyakov loop,
without affecting qualitatively the breaking 
of chiral symmetry. Only at much larger temperature chiral symmetry becomes
restored.

In order to understand better the role of the center degrees of freedom 
for confinement, 
in a series of papers \cite{Pepe} -- \cite{Liptak} lattice QCD has been 
studied for the gauge group G$_2$ where the center is trivial, i.e., consists
of only the identity element. These papers were all motivated by 
understanding various aspects of confinement. This is due to the fact that
G$_2$ Yang-Mills theory has a place in between Yang-Mills theory, adjoint QCD
and full QCD: It exhibits a linear rising potential with 
Casimir scaling \cite{Liptak}, which,
however, flattens out at large distances \cite{Pepe}, 
as is the case in full QCD. Its bound state
spectrum resembles adjoint QCD, as gluons can screen quarks and thus permit
color-neutral quark-gluon bound states in addition to 
glueballs and hadrons \cite{Pepe}.
Finally, it exhibits a first order phase transition, as does SU($N$) Yang-Mills
theory \cite{Pepe:2006er,Greensite,Cossu:2007dk}. Furthermore, gluonic
correlators seem not to differ qualitatively from the SU($N$) case \cite{Maas}.

So far, however, essentially 
nothing is known about aspects of chiral symmetry and its breaking for a
center-trivial gauge group. In particular, due to its intermediate status
between the other three types of theory, it is a-priori unclear whether
chiral symmetry breaking is present at all at low temperature, or, if chiral
symmetry is broken,
whether the chiral phase transition coincides with the thermodynamic one.

In our exploratory study we address for the quenched case the questions:

\begin{itemize}

\item
Is chiral symmetry broken in the confining phase? (Yes!)

\item 
Is chiral symmetry restored at high temperatures? (Yes!)

\item 
Does chiral symmetry restoration take place at the 
same temperature where the theory deconfines? (Yes!)

\item
How do the temporal boundary conditions of the Dirac operator
influence the spectral quantities and
thus observables for chiral symmetry?

\end{itemize}

These questions can be formulated and answered
in terms of spectral quantities of the Dirac
operator, in particular the density of eigenvalues near the origin or a 
possible spectral gap, which appears above the critical temperature $T_c$. 
In our paper we analyze spectral properties of the staggered lattice Dirac
operator using quenched G$_2$ gauge configurations below and above $T_c$.

\vskip10mm
\noindent
{\Large The role of temporal fermionic boundary conditions in SU($N$)}
\vskip5mm
\noindent 
When QCD at finite temperature is considered in the Euclidean path integral 
formalism the time extent of space-time is finite. In such a 
setting the temporal boundary conditions become a relevant issue. 
During recent years 
the role of the temporal fermion boundary conditions was analyzed 
in several lattice QCD studies \cite{EMIetal1} -- \cite{wipf}. 
A part of these studies
\cite{EMIetal1,gattringeretal,gaschae,Laplace,EMIetal2} 
was motivated by analyzing  
caloron and dyon signatures of the QCD vacuum, where for the case of calorons 
\cite{kvb} specific signatures of Dirac eigenmodes for different boundary 
conditions are known \cite{kvbzeromode}. 
Another motivation was a possible persistence of the
chiral condensate above $T_c$ 
\cite{Stephanov,Christ,z3paper,bornyakov,kovacs}, 
and more recently the temporal fermion
boundary conditions were used to relate observables for chiral symmetry to 
observables for center symmetry \cite{grazreburg,dualcond,soeldner,wipf}. 

Although the motivations of the studies \cite{EMIetal1} -- \cite{wipf} are 
different, one may infer two common observations 
which are of interest for the present paper:
\begin{enumerate}
\item
Below the critical temperature $T_c$ spectral quantities of the 
Dirac operator are insensitive to the temporal boundary condition. 
\item
Above $T_c$ spectral quantities feel the boundary conditions, 
but only the boundary angle relative to the phase of the Polyakov loop is 
relevant. 
\end{enumerate}
 
Let us be a little bit more explicit on the second observation. The temporal 
fermionic boundary condition may be parameterized by an angle 
$\varphi \in [-\pi,\pi)$ such that it reads
\begin{equation}
\psi(\vec{x},\beta) \; = \; e^{i \varphi} \, \psi(\vec{x},0) \; .
\label{bc}
\end{equation}
In this equation $\beta$ denotes the extent of the Euclidean time direction,
i.e., the inverse temperature (in lattice units). 
It is obvious, that the choice $\varphi = \pi$
gives rise to the usual anti-periodic boundary conditions. Here, however, we
allow for more general boundary conditions and use the 
angle $\varphi$ as a free parameter to probe the system. 

\begin{figure}[t]
\begin{center}
\includegraphics[height=60mm,clip]{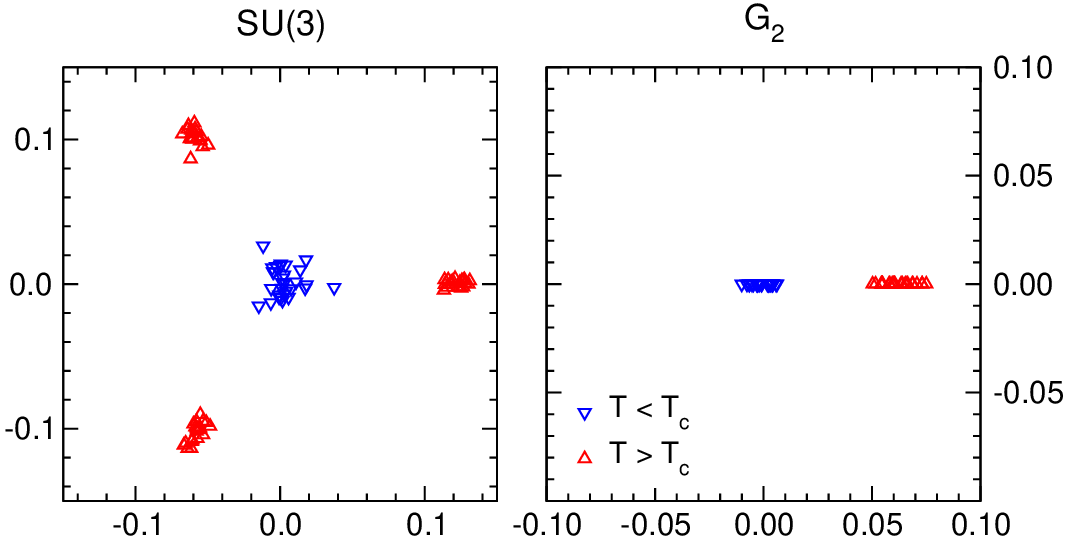}
\end{center}
\caption{Scatter plots of the Polyakov loop $P$
in the complex plane for gauge configurations below and above $T_c$. The
lhs.\ plot is for gauge group SU(3), while the rhs.~shows the case of G$_2$.}
\label{ploopscatter}
\end{figure}

The second relevant angle in the above listed observation 
is the phase $\theta_P$ of the (space-averaged) Polyakov loop
\begin{equation}
P \; = \; \frac{1}{V_3} \sum_{\vec{x}} \mbox{Tr} \, \prod_{x_4 = 1}^\beta 
 U_4(\vec{x},x_4) \; .
\end{equation}
In the quenched theory below $T_c$ the expectation value of $P$ vanishes,
while above the transition a non-vanishing expectation value emerges. 
For the case of the gauge group SU($N$), the values for the Polyakov loop
phases $\theta_P$ scatter around the phases of the center 
$\mathds{Z}_N$ of SU($N$), i.e., $\theta_P \sim n 2\pi/N, n = 0,1 ...,\, N-1$. 
Only for infinite volume the underlying center symmetry of the theory becomes 
broken and the Polyakov loop assumes a fixed phase $\theta_P$ of one of the
center values. These properties of $P$ are illustrated in the lhs.\ plot of
Fig.~\ref{ploopscatter} for the case of SU(3), where we show a scatter 
plot of the Polyakov loop values in 
the complex plane for two ensembles below and
above $T_c$. It is obvious that above $T_c$ the phases scatter around the
values $0, 2\pi/3$ and $4\pi/3$.  The subsets of configurations where the
Polyakov loop assumes a single one of the possible phases will be referred to
as {\sl Polyakov loop sectors}. In particular we will call the set 
of configurations
where the Polyakov loop is essentially real, i.e., $\theta_P \sim 0$, 
the {\sl real Polyakov loop sector}.

The second observation from the list  
above can now be formulated precisely: Only the sum
$\sigma$ of the boundary angle $\varphi$ and the phase $\theta_P$ of the
Polyakov loop, 
\begin{equation}
\sigma \; = \; \varphi \, + \, \theta_P \;\; \mbox{mod} \, 2\pi \; , 
\label{sigmadef}
\end{equation}
is relevant for spectral observables of the Dirac operator. 
In other words, the change of observables found when switching from one 
center sector of the Polyakov loop to another one
can be compensated by shifting the fermionic boundary conditions. 
This has been observed \cite{EMIetal1} -- \cite{wipf} in the gauge groups
SU(2) and SU(3) 
for a wide range of quantities, ranging from the
spectral gap to localization properties of zero and near-zero modes.  

In its most compact form the results for the spectrum 
may be written down as a generalization 
\cite{dualcond} of the well known Banks-Casher formula \cite{baca}. This
formula relates the chiral condensate 
$\langle \overline{\psi} \psi \rangle$ to the
density $\rho(0)$ of the Dirac eigenvalues at the origin,
\begin{equation}
\langle \overline{\psi} \psi \rangle_\sigma \; = \; - \pi \, \rho(0)_\sigma \; .
\label{bacageneral}
\end{equation}
We attach the subscript $\sigma$ to denote the total angle, consisting of  
the Polyakov loop phase and the boundary angle used for the evaluation of
the two sides. 

Below $T_c$ the spectrum extends all the way to the origin and a non-vanishing 
$\rho(0)$ exists which is independent of $\sigma$. Thus 
$\langle\overline{\psi} \psi \rangle \neq 0$ and chiral symmetry is
broken. Above $T_c$ a gap is expected to open up in the spectrum such that 
$\rho(0)$ vanishes and chiral symmetry is restored 
($\langle\overline{\psi} \psi \rangle = 0$).

The interesting observation is that for a zero total angle $\sigma = 0$ 
a non-vanishing density of eigenvalues $\rho(0)$ and thus a non-zero 
chiral condensate persists also above $T_c$. Such a zero total angle 
$\sigma = 0$ is, e.g., obtained when periodic 
boundary conditions ($\varphi = 0$) are used in the real Polyakov loop 
sector ($\theta_P = 0$). Another possibility for $\sigma = 0$ 
would be anti-periodic boundary
conditions for a Polyakov loop with phase $\theta_P = \pi$, 
which is possible for gauge
group SU($N$) with even $N$. More generally it may be shown \cite{gaschae}
that the spectral gap has a sine-like dependence on $\sigma$
and thus closes completely at $\sigma = 0$.
  
Having identified the role of the center $\mathds{Z}_N$ of the gauge groups 
SU($N$) and the corresponding phase sectors of the Polyakov loop for 
the spectral 
quantities, an interesting question comes up: How do spectral quantities, and
thus the chiral condensate below
and above $T_c$, behave when the gauge group has a trivial center, i.e.,
the center consists of only the identity element, as is the case for 
the group G$_2$? And how do the fermionic
temporal boundary conditions influence the Dirac spectrum 
in this case?

Before we address these questions in the next sections, we conclude with
remarking that also for the case of G$_2$ the finite temperature 
transition is signaled 
by a changing expectation value of the Polyakov loop \cite{Pepe:2006er}. 
Below $T_c$ this
expectation value is zero, while at $T_c$ it jumps in a first order 
transition\footnote{This transition should not be confused with the bulk
transition \cite{Pepe:2006er,Greensite} at a lower inverse coupling, 
$\beta = 9.45$, which 
is, however, irrelevant for all
considerations here.} to a non-vanishing value. This behavior is illustrated
in the rhs.\ plot of Fig.~\ref{ploopscatter} 
(see also Fig.~\ref{ploop} below), 
where we again show scatter
plots of the Polyakov loop values below and above $T_c$, now for gauge group 
G$_2$. What is immediately obvious is the fact that the Polyakov values are
real, due to the existence of real representations of G$_2$.

Furthermore, above $T_c$ no non-trivial phase structure appears. However,
in the infinite volume limit the Polyakov loop
in both phases will necessarily vanish, since there is 
no asymptotic string tension \cite{Pepe:2006er}. The phase
transition is nonetheless physical, as it is manifest also in the free energy 
\cite{Pepe:2006er,Cossu:2007dk}.

\vskip10mm
\noindent
{\Large The setting of our calculation}
\vskip5mm
\noindent
{\large Technicalities}
\vskip3mm
\noindent

For our simulations we used the standard Wilson action with the links in
a fundamental, but complex, 7-dimensional 
representation \cite{Pepe:2006er,Greensite}.

For our analysis we use quenched G$_2$ configurations at finite temperatures
below and above the critical temperature. We worked with different lattice
sizes ranging from $8^3 \times 4$ to $14^3 \times 6$, and $\beta$ 
values between 9.45 and 10.
All results we show are for lattice size $12^3 \times 6$. The
configurations are generated with a hybrid heat-bath 
\cite{Pepe:2006er,Greensite} 
and overrelaxation \cite{Maas} algorithm. Details can be found in \cite{Maas}.
We always made several independent runs to reduce autocorrelation effects, 
with no more than ten configurations per run generated for 
fermionic observables. We
allowed for 28 to 34 thermalization sweeps and between 80 to 140 decorrelation
sweeps between consecutive measurements.

For scale setting we use the string tension determined from Wilson
loops\footnote{We thank Ludovit Liptak for providing us with his  results,
  partly unpublished, of a high precision determination of the scale.}
\cite{Liptak}.  In some of our plots we use units of GeV for illustration
purposes. These were introduced by using for the G$_2$ string tension the
value known for SU(3) ($\sigma = (0.44$ GeV$)^2$).  Note that since even
quenched G$_2$ Yang-Mills theory does not exhibit an asymptotic linear rising
potential, the intermediate distance string-tension has been used to fix the
scale. Intermediate distance is here the distance where Casimir scaling of the
string tension is observed \cite{Liptak}. At larger distances the potential
flattens out in quenched G$_2$, while it becomes $N$-ality scaling in SU($N$)
Yang-Mills theory.  Note that this is nonetheless equivalent to the procedure
to set the scale in full QCD, as also there in the
Casimir-region the scale is fixed.
The critical coupling $\beta_c$, and thus temperature $T_c$, was taken from
\cite{Cossu:2007dk}, but we also observed them by Polyakov-loop and plaquette
observables, reproducing the results of \cite{Cossu:2007dk}.  

The fermionic observables computed from complete Dirac operator spectra were
evaluated on typically 40 configurations at each temperature.   The error
bars we show are statistical errors determined from single elimination
Jackknife. 

For the analysis of the fermionic variables we use the staggered lattice Dirac
operator
\begin{equation}
D_{xy} \; = \; 
\sum_{\mu = 1}^4 \! \frac{\eta_\mu(x)}{2a} 
\Big[ U_\mu(x) \delta_{x+\hat{\mu},y} - 
U_\mu(x\!-\!\hat{\mu})^\dagger \delta_{x-\hat{\mu},y} \Big],
\label{staggeredD}
\end{equation}
with the staggered sign function   
{$\eta_\mu(x)\!= \!(-1)^{x_1+  \, ... \, + x_{\mu-1}}$}. 
The coordinates $x,y$ run over all sites of the 4-di\-men\-sio\-nal $L^3
\times N_4$ lattice. 
The gauge link variables $U_\mu(x)$ are
elements  of the gauge group G$_2$. The staggered Dirac operator is an
anti-Hermitian matrix and has its eigenvalues on the imaginary axis. 
We evaluate complete spectra using a parallel implementation of
standard LAPACK routines. For each ensemble the complete spectra where
computed for several different fermionic temporal boundary conditions
(\ref{bc}). These are most simply implemented by attaching the phase to the
last temporal link of the lattice. All the fermionic observables which we
discuss below (eigenvalue density, chiral condensate, spectral gap and the
dual chiral condensate) may be computed from the spectra at the different
boundary conditions.  

\begin{figure}[t]
\begin{center}
\includegraphics[height=60mm,clip]{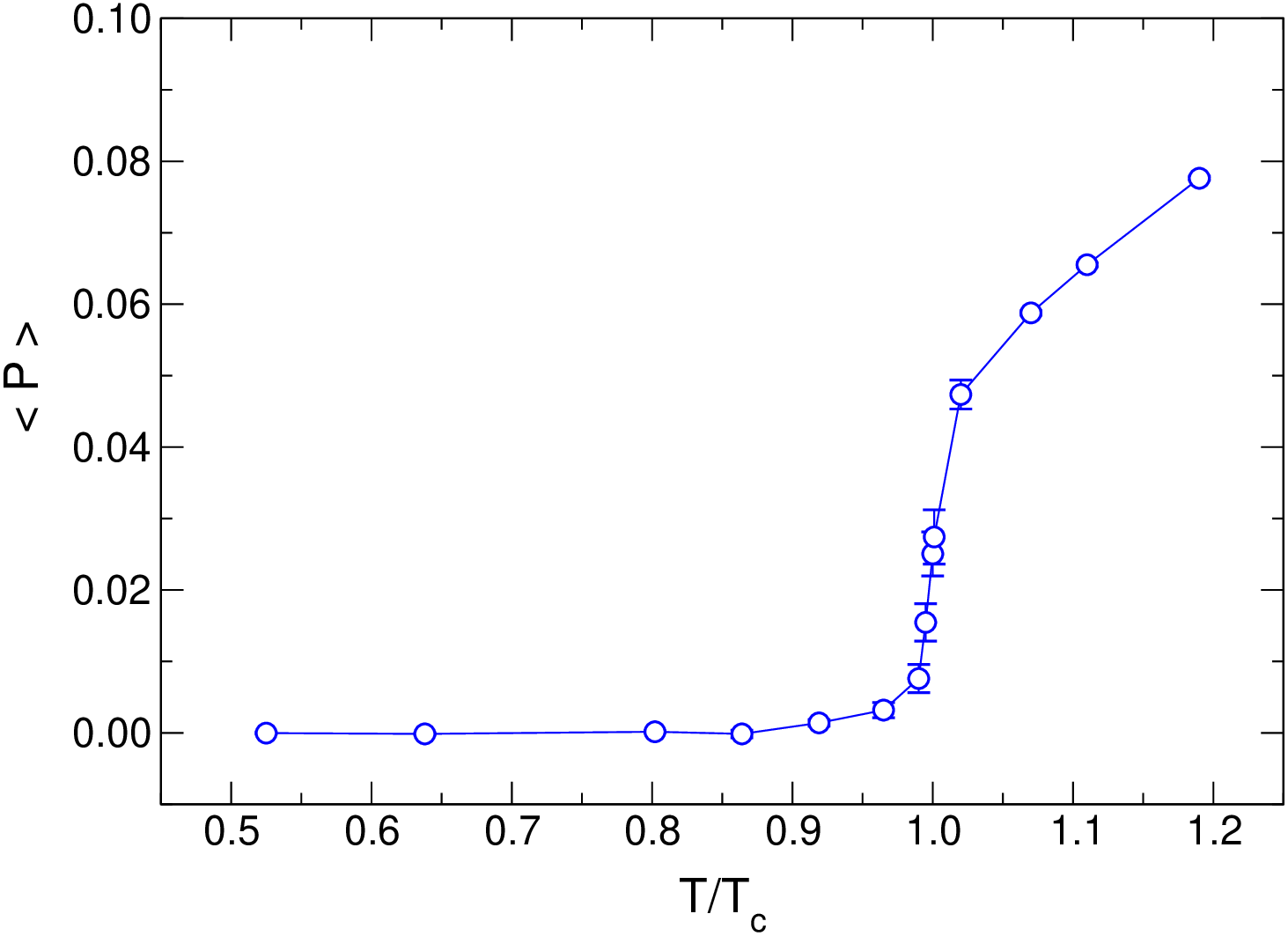} 
\end{center}
\vspace{-2mm}
\caption{Expectation value of the Polyakov loop as 
function of temperature.} 
\label{ploop}
\end{figure}

\newpage
\noindent
{\large Gluonic observables}
\vskip3mm
\noindent
For illustration purposes we briefly discuss the results for the Polyakov loop.
In particular the plot of the Polyakov loop as a function of the temperature
will later be useful for comparison when we present our results for fermionic
observables as a function of the temperature. 

In Fig.~\ref{ploop} we show our results for the Polyakov loop
as a function of the
temperature on a $12^3 \times 6$ lattice.  The plot clearly shows a steep rise
of the expectation value of the Polyakov loop at the critical temperature $T_c$.

\begin{figure}[h!]
\begin{center}
\includegraphics[height=55mm,clip]{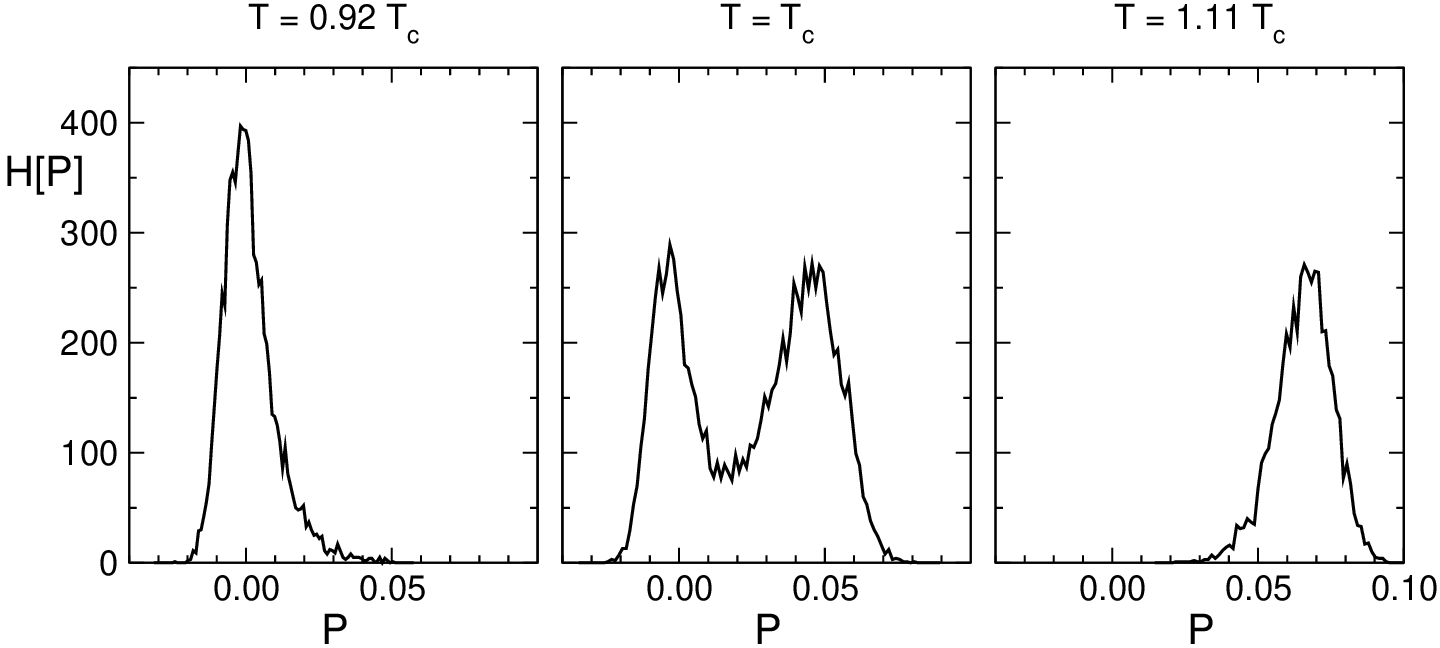} 
\end{center}
\vspace{-2mm}
\caption{Histograms for the values of the Polyakov loop $P$. We compare 
three temperatures below (lhs.\ plot), at (center) and above 
$T_c$ (rhs.\ plot). Similar observations have been made in \cite{Pepe:2006er}.} 
\label{ploophistograms}
\end{figure}

The transition at the critical temperature is known to be of first order
\cite{Pepe:2006er,Cossu:2007dk}. This is also clearly  seen in our ensembles
as is demonstrated in Fig.~\ref{ploophistograms}, where we show for the $12^3
\times 6$ lattice histograms of the Polyakov loop at  $T < T_c$ (lhs.~plot),
$T = T_c$ (center) and $T > T_c$ (rhs. plot).   The double peak structure in
the center plot clearly shows the coexistence of two phases at the critical
temperature, thus indicating the first order transition.  This result can also
be deduced from the free energy \cite{Cossu:2007dk}.

This similarity of gluonic observables for G$_2$ and SU($N$) gauge theories
was discussed previously in the context of confinement
\cite{Pepe,Maas:2005ym}, and it was conjectured that gluonic observables
should coincide qualitatively in both cases \cite{Maas:2005ym}.

\vskip3mm

\newpage
\noindent
{\Large Dirac spectra and fermionic observables}
\vskip5mm
\noindent
{\large Spectral density at the origin and the chiral condensate}

\vskip3mm
\noindent
As we have discussed above, the eigenvalue density of the Dirac operator near
the origin is related to the chiral condensate via the Banks-Casher relation
(\ref{bacageneral}). For the case of SU($N$) gauge theories we know that the
spectrum behaves differently for different boundary conditions. In
particular for the gauge ensemble restricted to configurations with Polyakov
loop in the real sector, the density  vanishes above $T_c$ for all boundary
conditions, except for the case where the fermionic boundary condition in time
direction is chosen periodic. For the other sectors of the Polyakov loop, the
condensate persists for accordingly shifted boundary angles 
$\varphi = 2\pi (N-1)/N,
\varphi = 2\pi (N-2)/N ...\;$.

\begin{figure}[t]
\begin{center}
\hspace*{-3mm}
\includegraphics[height=45mm,clip]{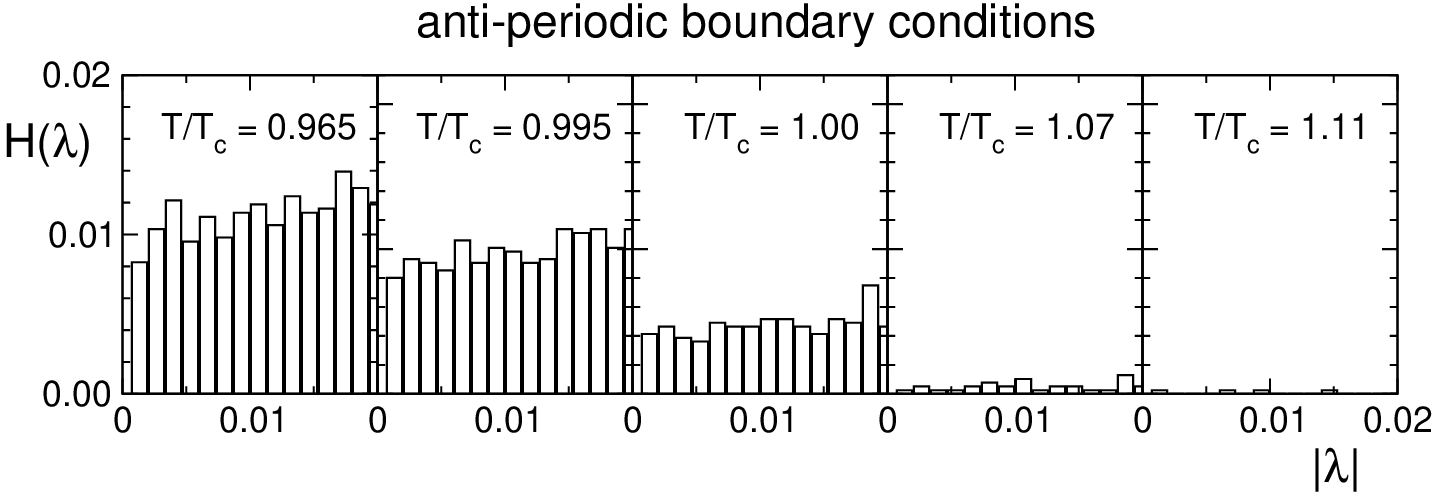} 
\vskip3mm
\hspace*{-3mm}
\includegraphics[height=45mm,clip]{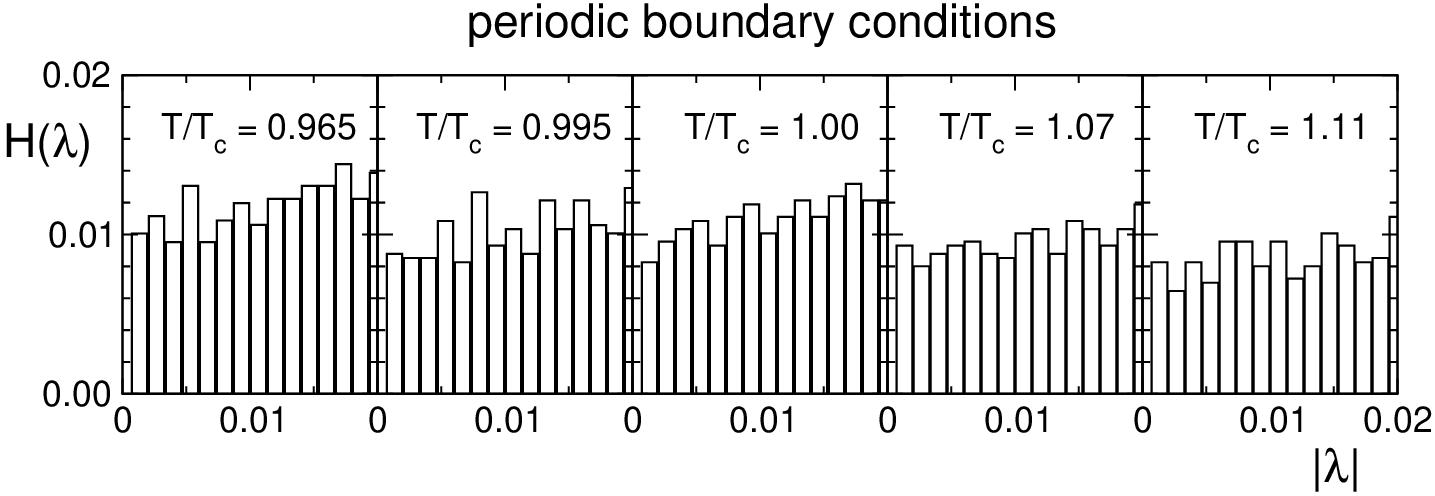} 
\end{center}
\caption{Distribution of the Dirac eigenvalues $\lambda$ as a function of 
$|\lambda|$. We show the distribution for several values of the temperature
  below and above $T_c$ and compare anti-periodic (top panel) to periodic
  (bottom) boundary conditions.}  
\label{evaldensity}
\end{figure}

For the gauge group G$_2$ the Polyakov loop is always real and in order to test
if this case is similar to SU($N$) we need to compare periodic and anti-periodic
temporal fermion boundary conditions. In Fig.~\ref{evaldensity} we show
histograms of the eigenvalues near the origin for several temperatures below and
above $T_c$. The top panel of plots is for anti-periodic boundary conditions,
while at the bottom we show the periodic case.

\begin{figure}[t]
\begin{center}
\includegraphics[height=60mm,clip]{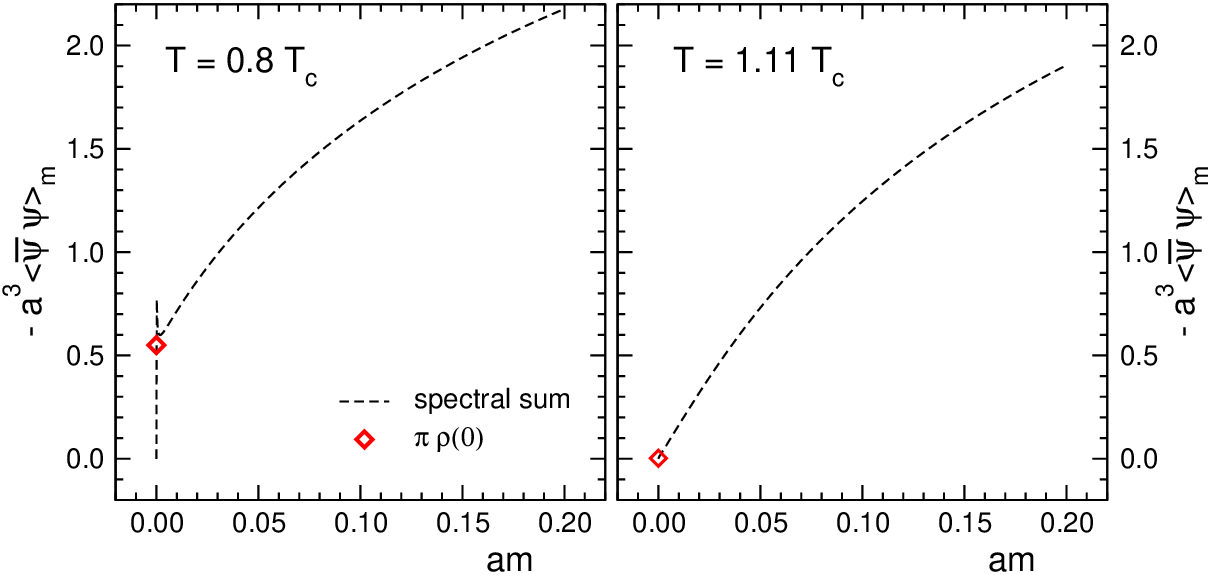} 
\end{center}
\caption{The condensate $\langle \overline{\psi} \psi
\rangle_m$ as a function of the quark mass (dashed curve) compared to the
result for the chiral condensate as obtained from the spectral density method
(symbols). All quantities are in lattice units and were obtained on our $12^3
\times 4$ configurations. We compare $T = 0.8 T_c$ 
(lhs.~plot) to $T = 1.11 T_c$ (rhs.~plot).}
\label{rawcondensate}
\end{figure}

For the anti-periodic case we find that below $T_c$ the eigenvalue density
extends all the way to the origin. At $T_c$ it starts to drop and vanishes above
the critical temperature. For the periodic boundary conditions the situation is
different, and the density at the origin survives also above $T_c$. The
histograms clearly demonstrate that the spectral density behaves similar to what
was found for SU($N$) when the real Polyakov loop sector is selected.

Since the spectral density at the origin and the chiral condensate  $\langle
\overline{\psi} \psi \rangle$ are linked through the Banks-Casher formula
(\ref{bacageneral}), it is natural to now study the condensate as a function of
the temperature for periodic and anti-periodic  boundary conditions. 

In our analysis of the condensate we compared two different ways for its
determination. First we computed $\langle \overline{\psi} \psi \rangle$ 
from the
density near the origin as determined from the histograms. Our second
determination was based on a direct evaluation of the condensate at a finite
fermion mass $m$, which is computed as a spectral sum of the Dirac eigenvalues
$\lambda_i$:
\begin{equation}
\langle \overline{\psi} \psi \rangle_m \; = \; - \frac{1}{V}
\sum_i \frac{1}{\lambda_i + m} \, .
\label{specsum}
\end{equation}
The proper chiral condensate is obtained by performing the limit $m \rightarrow
0$ after the infinite volume limit $V \rightarrow \infty$ is taken. On a finite
volume, as one is restricted to in a numerical analysis, the condensate must
vanish exactly, as no spontaneous symmetry breaking is possible in a finite
system. This is clearly seen in our data for $\langle \overline{\psi} \psi
\rangle_m$ which vanish for very small $m$. However, before vanishing
completely, $\langle \overline{\psi} \psi \rangle_m$ shows a long and pronounced
shoulder which may be extrapolated to vanishing mass. Comparing the results from
this extrapolation to the determination from the spectral density we always
found the discrepancies to be in the one percent range, showing that the two
methods give the same result.

We demonstrate this agreement explicitly in Fig.~\ref{rawcondensate}, where we show the
result from the spectral sum (\ref{specsum}) as a function of the mass
parameter (dashed curve) and compare it to the outcome from the spectral
density method (symbol). We show the results for $T = 0.8 T_c$ where the
condensate is finite (lhs.~plot) and for $T = 1.11 T_c$ (rhs.)
where the condensate
vanishes (using anti-periodic temporal boundary conditions in both cases). 
The lhs.\ plot shows the described  shoulder-type behavior, before it drops to
zero at vanishing mass. Just before this drop we observe a small spike which
is due to isolated small eigenmodes on individual configurations -- an
observation we made for some of our ensembles. When one ignores this spike and
extrapolates the shoulder to vanishing quark mass, one ends up exactly on 
the spectral density result indicated by the symbol. On the rhs.~plot no
shoulder is observed and the condensate directly extrapolates to zero in
agreement with the spectral method for that case.

\begin{figure}[t]
\begin{center}
\includegraphics[height=65mm,clip]{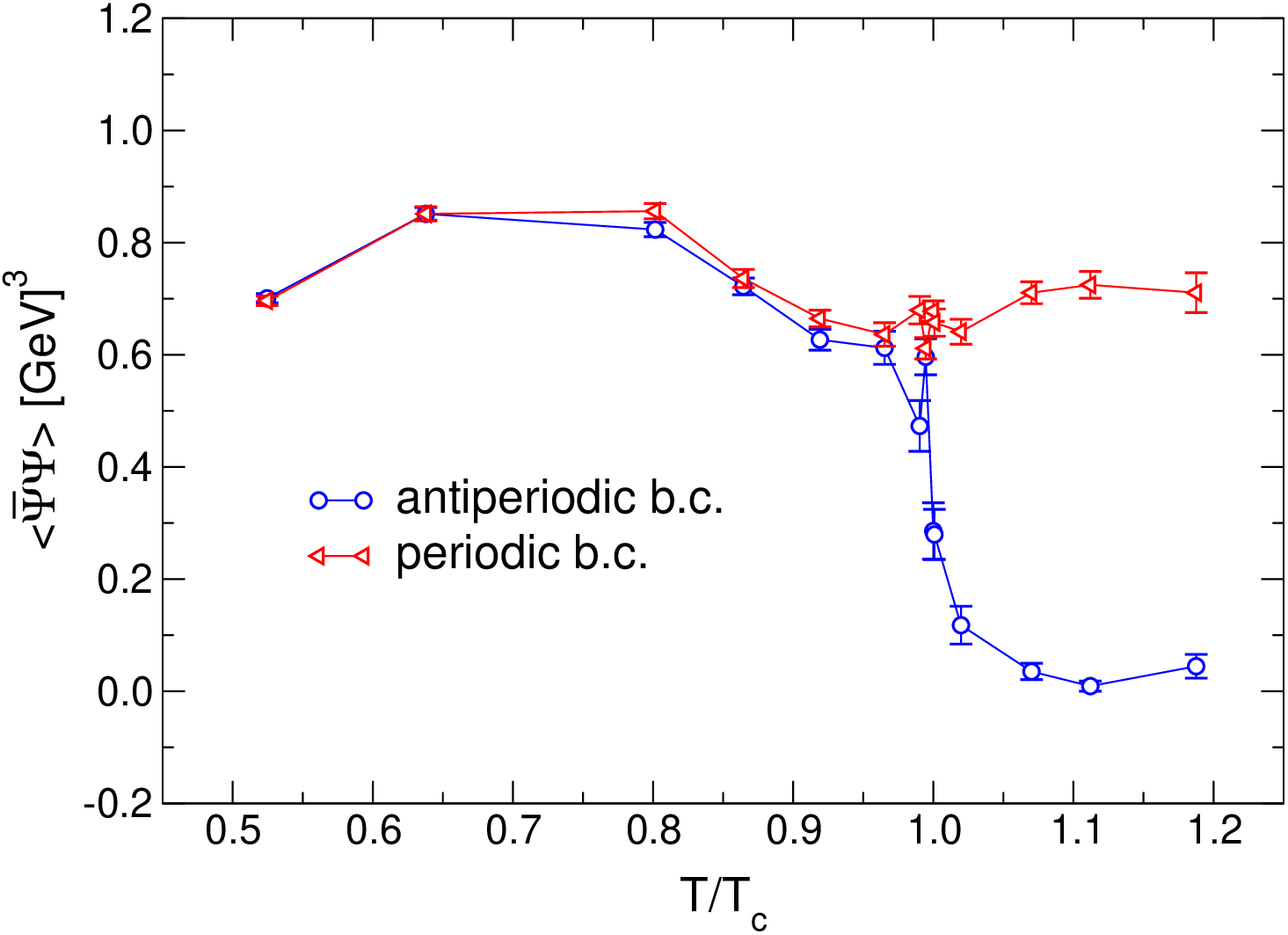} 
\end{center}
\caption{The chiral condensate as a function of the temperature. 
We compare the results for anti-periodic temporal boundary conditions to the
case of periodic boundary conditions.}  
\label{condensate}
\end{figure}

The results for the chiral condensate as a function of the temperature are shown
in Fig.~\ref{condensate}, where we again compare periodic and anti-periodic
boundary conditions. As was already suggested by the histograms in
Fig.~\ref{evaldensity}, only the anti-periodic case shows a restoration of 
chiral symmetry, i.e., a condensate that vanishes above $T_c$. For the periodic
case, where the boundary condition is in phase with the Polyakov loop, we see
that the condensate persists also above $T_c$. Again we find the same picture as
for SU($N$) gauge theory in the real Polyakov loop sector.

Let us finally stress that the fact that the condensate for anti-periodic
boundary conditions drops at the same critical temperature $T_c$ where also
the Polyakov loop and the free energy indicate the transition  (compare
Fig.~\ref{ploop}), is in itself a remarkable finding. One cannot a priori
expect that this is the case, as is known from the example of SU(3) 
with quarks in the
adjoint representation \cite{adjointtransition}. On the other hand it was
shown in \cite{grazreburg} -- \cite{wipf} that with the help of boundary
conditions the chiral condensate may be transformed into a generalized
Polyakov loop. This transformation suggests a strong link between chiral
symmetry and confinement, although the exception of adjoint quarks coupled to 
SU(3) gluons (and likely
for SU($N$)) still needs to be understood in this connection.  In
particular, it would be highly interesting to investigate whether for adjoint
G$_2$ quarks the transitions would also be distinct and thus would be a
general feature of the adjoint representation, or whether this is specific to
adjoint SU($N$) fermions.

\vskip5mm
\noindent
{\large The spectral gap}
\vskip3mm
\noindent
Having obtained the behavior of the chiral condensate let us now look at
the spectral gap, i.e., the average size $\langle | \lambda_{min} | \rangle$
of the smallest eigenvalue $\lambda_{min}$. Below $T_c$ the
density of eigenvalues extends all the way to the origin, which on a finite
lattice gives rise to only a microscopic gap, which is a finite size effect
that may be described by random matrix theory. At $T_c$ a macroscopic 
gap opens up in the
spectrum, such that the density at the origin and thus the condensate vanish. 
However, as discussed, for SU($N$) the size of the gap depends on the total
angle of Polyakov loop and boundary phase. 

\begin{figure}[t]
\begin{center}
\includegraphics[height=65mm,clip]{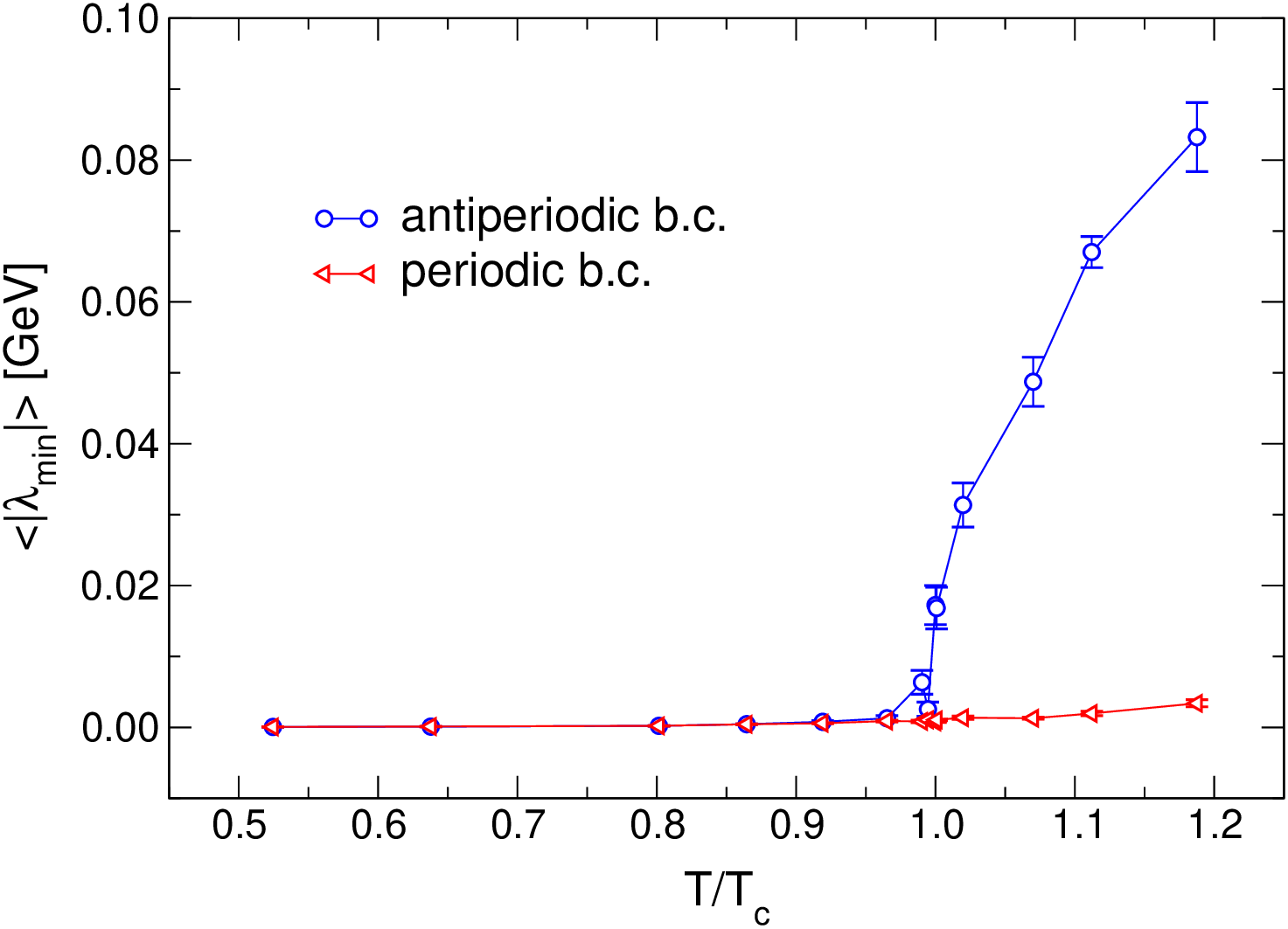} 
\end{center}
\caption{The spectral gap as a function of the temperature. 
We compare the results for anti-periodic temporal boundary conditions to the
case of periodic boundary conditions.}  
\label{gap}
\end{figure}

In Fig.~\ref{gap} we now analyze the spectral gap, i.e., 
$\langle | \lambda_{min} | \rangle$, as a function of the
temperature, comparing periodic and anti-periodic boundary conditions. 
The plot demonstrates very clearly that the gap opens up only when the
anti-periodic boundary conditions are used. For the periodic 
case, it remains closed. Since the Polyakov loop is real, the vanishing gap for
periodic boundary conditions obeys the $\sigma = 0$ criterion of 
Eq.~(\ref{sigmadef}). Thus also in this respect the gap behaves like in the
case of SU($N$). 

\begin{figure}[t]
\begin{center}
\includegraphics[height=65mm,clip]{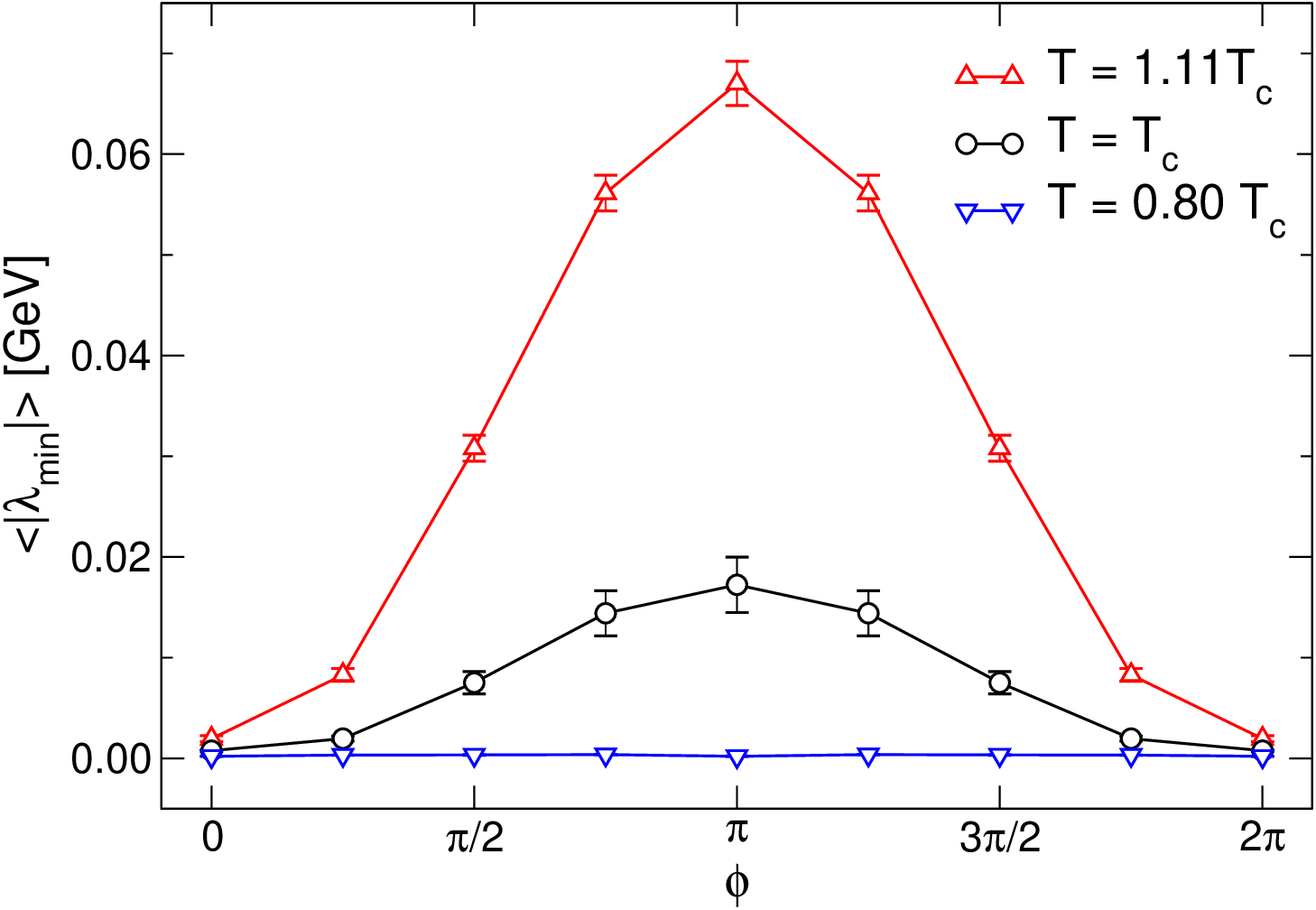}
\end{center}
\caption{The spectral gap as a function of the boundary angle $\varphi$. 
We compare the data for three different temperatures.}
\label{gap_vs_phi}
\end{figure}

To complete the analysis of the spectral gap, in Fig.~\ref{gap_vs_phi}  we
plot it as a function of the boundary angle $\varphi$, comparing three
temperatures. Below $T_c$ the gap is essentially zero, and the remaining
microscopic gap shows no $\varphi$-dependence within error bars. At $T_c$ a
slight sine-like behavior becomes visible, which becomes considerably more
pronounced above $T_c$. Using anti-periodic boundary conditions, i.e.,
$\varphi = \pi$, one picks up the value of the gap at maximal opening.  As
is seen in Fig.~\ref{gap}, this maximal gap grows monotonically over the range
of temperatures which we studied above $T_c$. For periodic boundary
conditions, i.e., $\varphi = 0 \equiv 2\pi$, the gap is closed. We stress that
this closing is not just visible for the smallest eigenvalue which defines the
gap, but as is obvious from the histograms in Fig.~\ref{evaldensity}, also the
higher eigenvalues come closer to zero. Thus indeed a finite spectral
density and thus
a non-vanishing condensate are found above $T_c$ for periodic boundary
conditions, as was already demonstrated in Fig.~\ref{condensate}.

\newpage
\noindent
{\large The dual chiral condensate}
\vskip3mm
\noindent

We finally come to an observable which is based on the non-trivial dependence
of spectral quantities above $T_c$ on the fermionic boundary condition, the
recently proposed \cite{dualcond} dual chiral condensate $\Sigma_1$. It is
obtained as a Fourier transform of the usual chiral condensate  with respect
to the boundary angle, 
\begin{equation}
\Sigma_1 \; = \; - \frac{1}{2\pi} \int_{-\pi}^\pi \!\!d\varphi \, e^{-i \varphi} 
\langle \overline{\psi} \, \psi \rangle_\varphi \; = \; 
\frac{1}{2 \pi} \int_{-\pi}^\pi \!\!d\varphi \, e^{-i \varphi} 
\frac{1}{V} \sum_{i} \frac{1}{m + \lambda^{(i)}_\varphi} \; .
\label{dualconddef}
\end{equation}
In the second step of this equation we have written the chiral condensate  as
a spectral sum over all eigenvalues $\lambda^{(i)}_\varphi$ of the lattice
Dirac operator, computed for boundary angle $\varphi$. The mass term is still
displayed in this sum, which may be sent to zero after the infinite volume
limit is taken. On a finite lattice one of the procedures which we have
discussed for the usual chiral condensate has to be applied.

The original motivation for the dual chiral condensate was the idea of
constructing an observable which connects the chiral condensate with
properties of the Polyakov loop, which in pure SU($N$) gauge theory is an
order parameter for the breaking of the center symmetry. The Fourier transform
with respect to the boundary angle selects from the closed 
fermion loops which the chiral condensate consists of, 
those which have a winding number of one. It is
obvious that these loops must transform under center transformations in the
same way as the Polyakov loop.  An important advantage of the dressed Polyakov
loop over the single thin Polyakov loop are its simple renormalization
properties which are inherited from the chiral condensate. The Fourier
transform allows one to switch between an observable for chiral symmetry
breaking to an observable for center symmetry with a simple renormalization.

\begin{figure}[t]
\begin{center}
\includegraphics[height=65mm,clip]{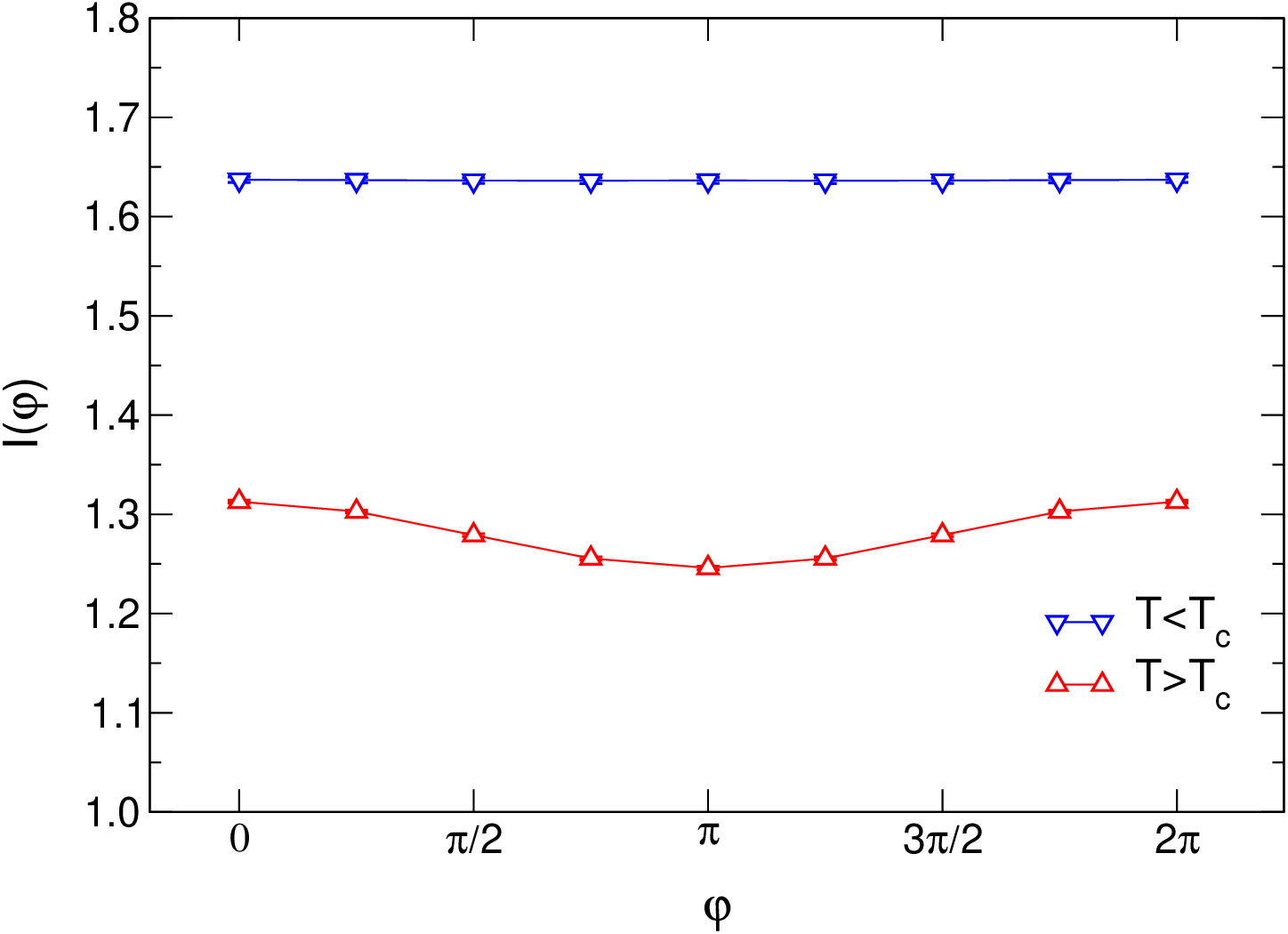}
\end{center}
\caption{The integrand of the dual chiral condensate at $am = 0.1$
as a function of the boundary angle $\varphi$. 
We compare the situation below and above $T_c$.}
\label{integrand_vs_phi}
\end{figure}

It is obvious from the definition (\ref{dualconddef}) that the dual chiral
condensate can have a non-vanishing value only when the integrand  
$I(\varphi) = V^{-1} \sum_i ( m + \lambda_\varphi^{(i)})^{-1}$ 
has a non-trivial 
dependence on the boundary angle $\varphi$. In Fig.~\ref{integrand_vs_phi} we 
show the integrand $I(\varphi)$ 
below and above $T_c$. Below $T_c$ it is independent of
$\varphi$ such that we expect a vanishing $\Sigma_1$ in the deconfined phase. 
Above $T_c$ a cosine-like behavior is seen, and the dual chiral condensate is
essentially the amplitude of this cosine. It is important to stress that the
modulation of $I(\varphi)$ is not due to the movement of a single eigenvalue,
but due to a collective response of the IR part of the spectrum to the
changing boundary angle $\varphi$ \cite{dualcond,wipf}. 

\begin{figure}[t]
\begin{center}
\includegraphics[height=65mm,clip]{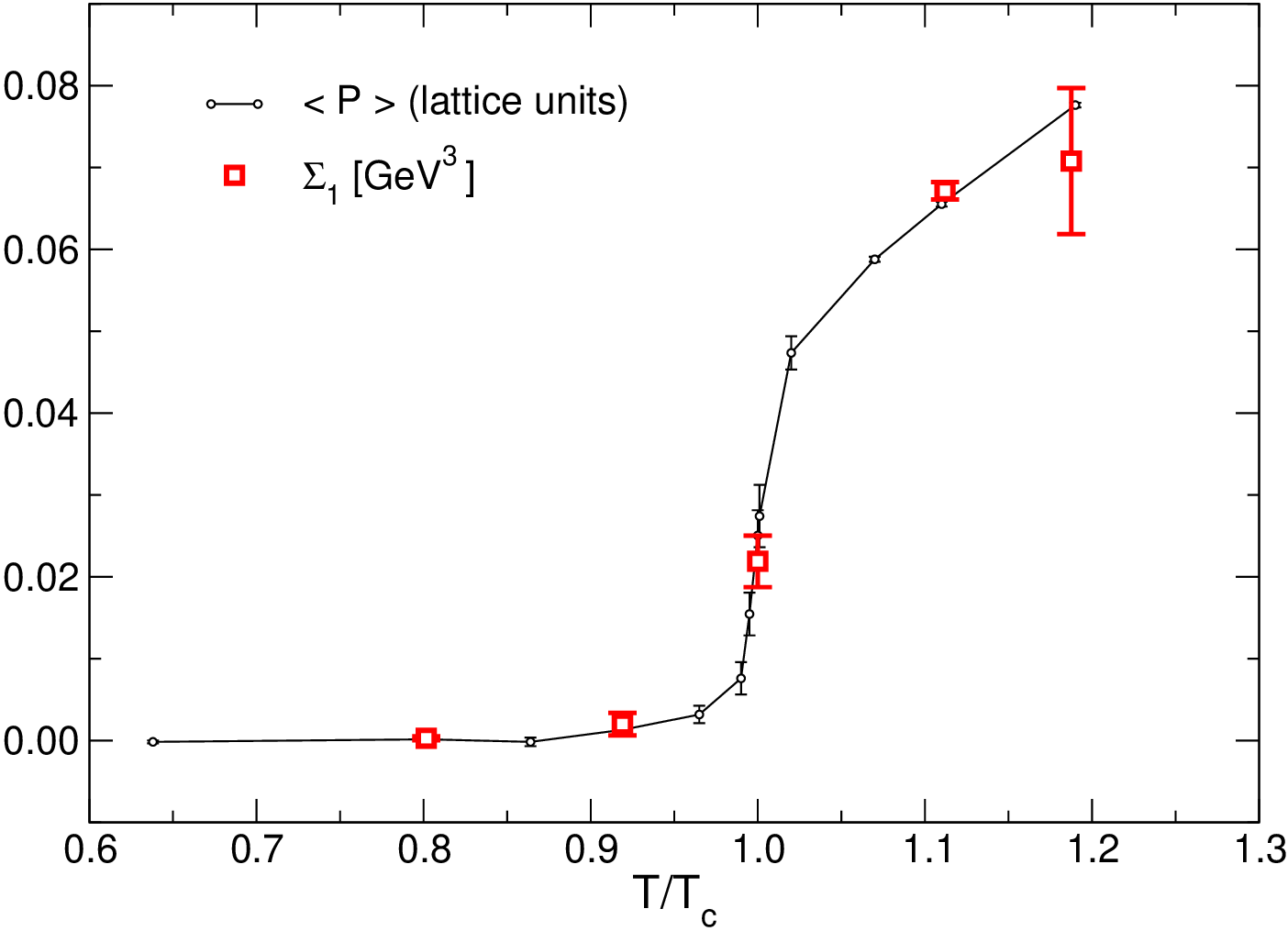}
\end{center}
\caption{The dual chiral condensate at $am = 0.1$
(large squares) and the expectation value 
of the thin Polyakov loop (small circles) as a function of the temperature.}
\label{dualcondplot}
\end{figure}

In Fig.~\ref{dualcondplot} we show the results for the dual chiral condensate
in physical units and compare it to the behavior of the thin
Polyakov loop in lattice units. In the case of a group with non-trivial center 
both these observables would test for the breaking of center
symmetry. Thus in this case qualitatively they should behave the 
same, i.e., vanish below
$T_c$ and have a finite value above $T_c$. 

The same behavior is observed also for the case of the center-trivial group 
$G_2$, as is obvious from the figure\footnote{We stress that the fact that the
  data for $\Sigma_1$ fall on top of the Polyakov loop values is a mere
  coincidence, since the latter are given 
in lattice units, and also are subject to
large renormalization effects.}. For $G_2$ there is however no simple
interpretation in terms of the breaking of the center symmetry - as is the
case for the Polyakov loop. It is again remarkable that also $\Sigma_1$,
which tests for the collective behavior of the IR spectrum as a function of
boundary conditions, behaves in the same way as one finds for SU(3).  

\newpage
\noindent
{\Large Concluding remarks}
\vskip5mm
\noindent
In this article we have analyzed the chiral condensate and 
spectral properties of the lattice Dirac operator for the center-trivial
gauge group G$_2$. The study was conducted for quenched gauge configurations
at various temperatures below and above $T_c$. 
Variating temporal fermionic boundary conditions were used to
probe the system. Of particular interest were the behavior of the chiral
condensate and the spectral gap above $T_c$ for various 
boundary conditions.

We have demonstrated that chiral symmetry is indeed spontaneously broken.
Furthermore, using anti-periodic boundary conditions one finds that the chiral
condensate vanishes at exactly the same temperature $T_c$ where also a
thermodynamic transition in the free energy is observed, and which also leaves
its trace in the Polyakov loop. In this respect the G$_2$ gauge theory behaves
in the same way as full QCD with fundamental quarks. 
In addition, we showed that at
$T_c$ a gap opens up in the spectrum as expected from the Banks-Casher
formula. 

As one switches to periodic boundary conditions the picture changes. The
chiral condensate remains finite above $T_c$ and no gap appears. This is the
same behavior as is found for gauge group SU($N$) if the sector with real
Polyakov loop is chosen. The other sectors show the same behavior after the
fermion boundary conditions are transformed with a center element of SU($N$). 
Above $T_c$ the spectral gap shows the characteristic sine-type 
dependence on the boundary angle known from SU($N$). 
Finally also the recently proposed dual chiral condensate, which is 
obtained through a Fourier transformation of the usual condensate with 
respect to the fermion boundary condition, shows the same behavior as 
expected from SU($N$). 

Thus we have found that for all spectral and chiral
observables which we inspected, the case of
gauge group G$_2$ behaves in exactly the same way as the gauge group 
SU($N$) when the sector with real Polyakov loop is selected. This is a natural
outcome, since the Polyakov loop is always real in G$_2$. These results
further support the picture that for chiral properties of a theory
the sector of the Polyakov loop only acts as a background field with a rather
trivial role: For G$_2$ only a single sector exists and for SU($N$) the
results from the different sectors may be mapped onto each other by a suitable
transformation of the fermion boundary conditions.

An interesting open puzzle concerning the underlying microscopic mechanism 
remains: In several papers \cite{chirsimcenter} it
was argued that center vortices play a non-trivial role also for the 
breaking of chiral symmetry. Our finding that in many respects the 
center-trivial gauge group G$_2$ produces the same chiral pattern as the 
real Polyakov loop sector of SU($N$)
thus should be understood also from a microscopic 
point of view. In this respect the proposed domain structure of the G$_2$ vacuum
may be of relevance \cite{Greensite}. This would imply that not the center
charge plays a significant role for infrared dynamics, but only the
existence of long range structures, as has already been conjectured
for gluonic properties \cite{Maas}.

\vskip6mm
\noindent
{\Large Acknowledgments}
\vskip3mm
\noindent
We thank Erek Bilgici, Falk Bruckmann, 
Christian Hagen, Tamas Kovacs, 
Christian Lang, Ludovit Liptak, Michael Ilgenfritz,
{\v{S}}tefan Olejn\'ik, and Uwe Wiese 
for discussions. The numerical analysis was
done at the ZID, University of Graz. J.~Danzer
is supported by the FWF Doktoratskolleg {\sl Hadrons in Vacuum, Nuclei and
Stars} (DK W1203-N08) and C.~Gattringer and 
A.~Maas by the FWF Project P20330.


\begin{thebibliography}{1234567}

\bibitem{greensite}
  J.~Greensite,
  Prog.\ Part.\ Nucl.\ Phys.\  {\bf 51} (2003) 1;
%
  M.~Engelhardt,
  Nucl.\ Phys.\ Proc.\ Suppl.\  {\bf 140} (2005) 92.


\bibitem{chirsimcenter}
  P.~de Forcrand and M.~D'Elia,
  Phys.\ Rev.\ Lett.\  {\bf 82} (1999) 4582;
%
  J.~Gattnar, C.~Gattringer, K.~Langfeld, H.~Reinhardt, A.~Sch\"afer, 
  S.~Solbrig and T.~Tok,
  Nucl.\ Phys.\  B {\bf 716} (2005) 105;
%
  PoS {\bf LAT2005} (2006) 301;
%
  R.~H\"ollwieser, M.~Faber, J.~Greensite, U.M.~Heller and {\v{S}}.~Olejn\'ik,
  arXiv:0805.1846 [hep-lat].
%
  V.G.~Bornyakov, E.M.~Ilgenfritz, B.V.~Martemyanov, S.M.~Morozov, 
  M.~M\"uller-Preussker and A.I.~Veselov,
  Phys.\ Rev.\  D {\bf 77} (2008) 074507;
%
  P.~Y.~Boyko {\it et al.},
  Nucl.\ Phys.\  B {\bf 756}, 71 (2006).

\bibitem{adjointtransition}
  F.~Karsch and M.~Lutgemeier,
  Nucl.\ Phys.\  B {\bf 550} (1999) 449;
%
  J.~Engels, S.~Holtmann and T.~Schulze,
  Nucl.\ Phys.\  B {\bf 724} (2005) 357.

\bibitem{Pepe}
  K.~Holland, P.~Minkowski, M.~Pepe and U.J.~Wiese,
  Nucl.\ Phys.\  B {\bf 668}, 207 (2003);
%
  Nucl.\ Phys.\ Proc.\ Suppl.\  {\bf 119}, 652 (2003);
%
%
  M.~Pepe,
  PoS {\bf LAT2005} (2006) 017
  [Nucl.\ Phys.\ Proc.\ Suppl.\  {\bf 153} (2006) 207].
%
%
  
\bibitem{Pepe:2006er}
  M.~Pepe and U.J.~Wiese,
  Nucl.\ Phys.\  B {\bf 768}, 21 (2007).

\bibitem{Greensite}
  J.~Greensite, K.~Langfeld, {\v{S}}.~Olejn\'ik, H.~Reinhardt and T.~Tok,
  Phys.\ Rev.\  D {\bf 75}, 034501 (2007).

\bibitem{Cossu:2007dk}
  G.~Cossu, M.~D'Elia, A.~Di Giacomo, B.~Lucini and C.~Pica,
  JHEP {\bf 0710} (2007) 100.

\bibitem{Maas}
  A.~Maas and {\v{S}}.~Olejn\'ik,
  JHEP {\bf 0802}, 070 (2008).

\bibitem{Liptak}
  L.~Liptak and {\v{S}}.~Olejn\'ik,
  arXiv:0807.1390 [hep-lat].

\bibitem{EMIetal1}
  E.M.~Ilgenfritz, B.V.~Martemyanov, M.~M\"uller-Preussker, S.~Shcheredin 
  and A.I.~Veselov,
  Phys.\ Rev.\  D {\bf 66} (2002) 074503;
%
  Nucl.\ Phys.\ Proc.\ Suppl.\  {\bf 119} (2003) 754;
%
  arXiv:hep-lat/0301008;
%
  C.~Gattringer {\it et al.},
  Nucl.\ Phys.\ Proc.\ Suppl.\  {\bf 129} (2004) 653;
%
  E.M.~Ilgenfritz, M.~M\"uller-Preussker and D.~Peschka,
  Phys.\ Rev.\  D {\bf 71} (2005) 116003.

\bibitem{gattringeretal}
  C.~Gattringer,
  Phys.\ Rev.\  D {\bf 67} (2003) 034507;
%
  C.~Gattringer and R.~Pullirsch,
  Phys.\ Rev.\  D {\bf 69} (2004) 094510;
%
  C.~Gattringer and S.~Solbrig,
  Nucl.\ Phys.\ Proc.\ Suppl.\  {\bf 152} (2006) 284.

\bibitem{gaschae}
  C.~Gattringer and S.~Schaefer,
  Nucl.\ Phys.\  B {\bf 654} (2003) 30.
%

\bibitem{Laplace}
  F.~Bruckmann and E.M.~Ilgenfritz,
  Phys.\ Rev.\  D {\bf 72} (2005) 114502;
%
  F.~Bruckmann and E.M.~Ilgenfritz,
  PoS {\bf LAT2005} (2006) 305.

\bibitem{EMIetal2}
  V.G.~Bornyakov, E.M.~Ilgenfritz, B.V.~Martemyanov and M.M\"uller-Preussker,
  arXiv:0809.2142 [hep-lat];
%
  V.G.~Bornyakov, E.M.~Ilgenfritz, B.V.~Martemyanov, S.M.~Morozov, 
  M.M\"uller-Preussker and A.I.~Veselov,
  Phys.\ Rev.\  D {\bf 76} (2007) 054505.

\bibitem{Stephanov}
  M.A.~Stephanov,
  Phys.\ Lett.\  B {\bf 375}, 249 (1996).

\bibitem{Christ}
  S.~Chandrasekharan and N.H.~Christ,
  Nucl.\ Phys.\ Proc.\ Suppl.\  {\bf 47} (1996) 527.


\bibitem{z3paper}
  C.~Gattringer, P.E.L.~Rakow, A.~Sch\"afer and W.~S\"oldner,
  Phys.\ Rev.\  D {\bf 66} (2002) 054502.

\bibitem{bornyakov}
  V.G.~Bornyakov, E.V.~Luschevskaya, S.M.~Morozov, M.I.~Polikarpov, 
  E.M.~Ilgenfritz and M.~M\"uller-Preussker,
  arXiv:0807.1980 [hep-lat];
%
  PoS {\bf LAT2007} (2007) 315.

\bibitem{kovacs}
  T.~Kovacs, PoS {\bf LAT2008} (2008) 198.

\bibitem{grazreburg}
  C.~Gattringer,
  Phys.\ Rev.\ Lett.\  {\bf 97} (2006) 032003;
%
  F.~Bruckmann, C.~Gattringer and C.~Hagen,
  Phys.\ Lett.\  B {\bf 647} (2007) 56;
%
  C.~Hagen, F.~Bruckmann, E.~Bilgici and C.~Gattringer,
  PoS {\bf LAT2007} (2007) 289;
%
  PoS {\bf LAT2008} (2008) 262
  [arXiv:0810.0899 [hep-lat]];
%
  E.~Bilgici and C.~Gattringer,
  JHEP {\bf 0805} (2008) 030.

\bibitem{dualcond}
  E.~Bilgici, F.~Bruckmann, C.~Gattringer and C.~Hagen,
  Phys.\ Rev.\  D {\bf 77} (2008) 094007.
%
\bibitem{soeldner}
  W.~S\"oldner,
  PoS {\bf LAT2007} (2007) 222.

\bibitem{wipf}
  F.~Synatschke, A.~Wipf and C.~Wozar,
  Phys.\ Rev.\  D {\bf 75} (2007) 114003;
%
  F.~Synatschke, A.~Wipf and K.~Langfeld,
  Phys.\ Rev.\  D {\bf 77} (2008) 114018.

\bibitem{kvb}
  T.C.\ Kraan and P.\ van Baal,
  Phys.\ Lett.\ B {\bf 428} (1998) 268;
%
  Nucl.\ Phys.\ B {\bf 533} (1998) 627;
%
  Phys.\ Lett.\ B {\bf 435} (1998) 389;
%
  F.~Bruckmann and P.~van Baal,
  Nucl.\ Phys.\ B {\bf 645} (2002) 105.

\bibitem{kvbzeromode}
  M.\ Garcia P\'erez, A.\ Gonz\'alez-Arroyo, C.\ Pena and P.\ van Baal,
  Phys.\ Rev.\ D {\bf 60} (1999) 031901;
%
  M.N.\ Chernodub, T.C.\ Kraan and P.\ van Baal,
  Nucl.\ Phys.\ Proc.\ Suppl.\ {\bf 83} (2000) 556.

\bibitem{baca}
T.~Banks and A.~Casher,
Nucl.\ Phys.\ B {\bf 169} (1980) 103.

\bibitem{Maas:2005ym}
  A.~Maas,
  Mod.\ Phys.\ Lett.\  A {\bf 20} (2005) 1797.

\end{thebibliography}
\end{document}